\documentclass{webofc}

\usepackage[varg]{txfonts}   
\usepackage{hyperref}
\usepackage{url}
\usepackage{siunitx}
\usepackage[version=4]{mhchem}
\usepackage{subcaption} 
\usepackage{wrapfig}

\newcommand{\DSnchan}{{2720 }}

\newcommand{\digmodel}{VX2745 }
\newcommand{\tpcdatarate}{\SI{2.5}{\giga \byte / \second} }

\newcommand{\fiducialvolume}{\SI{20}{t} }
\newcommand{\larvolume}{\SI{50}{t} }
\newcommand{\dcr}{\SI{40}{\hertz}}
\hypersetup{colorlinks=true,citecolor=blue,urlcolor=blue,linkcolor=blue}
\begin{document}
\title{Overview of the data acquisition system architecture for the DarkSide-20k experiment}
%
%

\author{\firstname{Maria Adriana,} \lastname{Sabia,}\inst{1,2}\fnsep\thanks{\email{maria.adriana.sabia@roma1.infn.it}} \\ 
        on behalf of the DarkSide-20k collaboration}


\institute{Physics Department, Sapienza Università di Roma, Roma 00185, Italy  
\and
           INFN Sezione di Roma, Roma 00185, Italy 
          }

\abstract{The DarkSide-20k (DS-20k) detector is designed to directly search for dark matter by detecting weakly interacting massive particles (WIMPs) scattering off nuclei in a 20 tonnes target of low radioactivity Argon from underground sources, in a dual-phase time projection chamber (TPC). The scintillation light produced by these interactions is recorded by custom silicon photomultiplier (SiPMs) assemblies comprising the detector’s \DSnchan readout channels.  

The data acquisition system (DAQ) for the DS-20k experiment operates in a triggerless mode, handling the estimated \SI{3}{\giga \byte/ \second} flow of digitized data from MC simulation from the entire detector, dominated by the TPC. Signals are digitized by 48 commercial \digmodel CAEN \(16\)-bit, \SI{125}{\mega S /\second}, high channel density waveform digitizers. This digitized data is then transferred to 24 Front-End Processor (FEP) machines for an initial data reduction stage. Subsequently, the processed data is sent to a separate set of Time Slice Processor (TSP) computers, where it is assembled into fixed-length time series for further analysis and storage for offline processing.  

}
\maketitle
\section{Introduction}
\label{sec:intro}
The nature of dark matter (DM) remains one of the most significant unresolved mysteries in contemporary physics. Numerous experiments have explored various hypotheses, but a definitive solution continues to elude researchers. Direct detection has emerged as a field of intense global interest among the diverse approaches. This method focuses on observing the exceedingly rare interactions of dark matter particles with regular matter. 

Liquid argon detectors have gained prominence among the many available technologies due to their exceptional pulse-shape discrimination based on the unique scintillation characteristics of argon \cite{PhysRevB.27.5279, BOULAY2006179, Adhikari_2021} and needed to distinguish electron recoils (ER) from nuclear recoils (NR). Based on that, the DarkSide-20k (DS-20k) experiment is committed to advancing the sensitivity of direct detection experiments to the level of the neutrino floor, a limit imposed by the unavoidable neutrino background inherent to current detection methods \cite{PhysRevLett.127.251802}.

\section{DarkSide-20k experiment}
\label{sec:ds20k}
\begin{figure*}[ht!]
\centering
\begin{subfigure}{0.45\textwidth} 
    \centering
    \includegraphics[width=\textwidth]{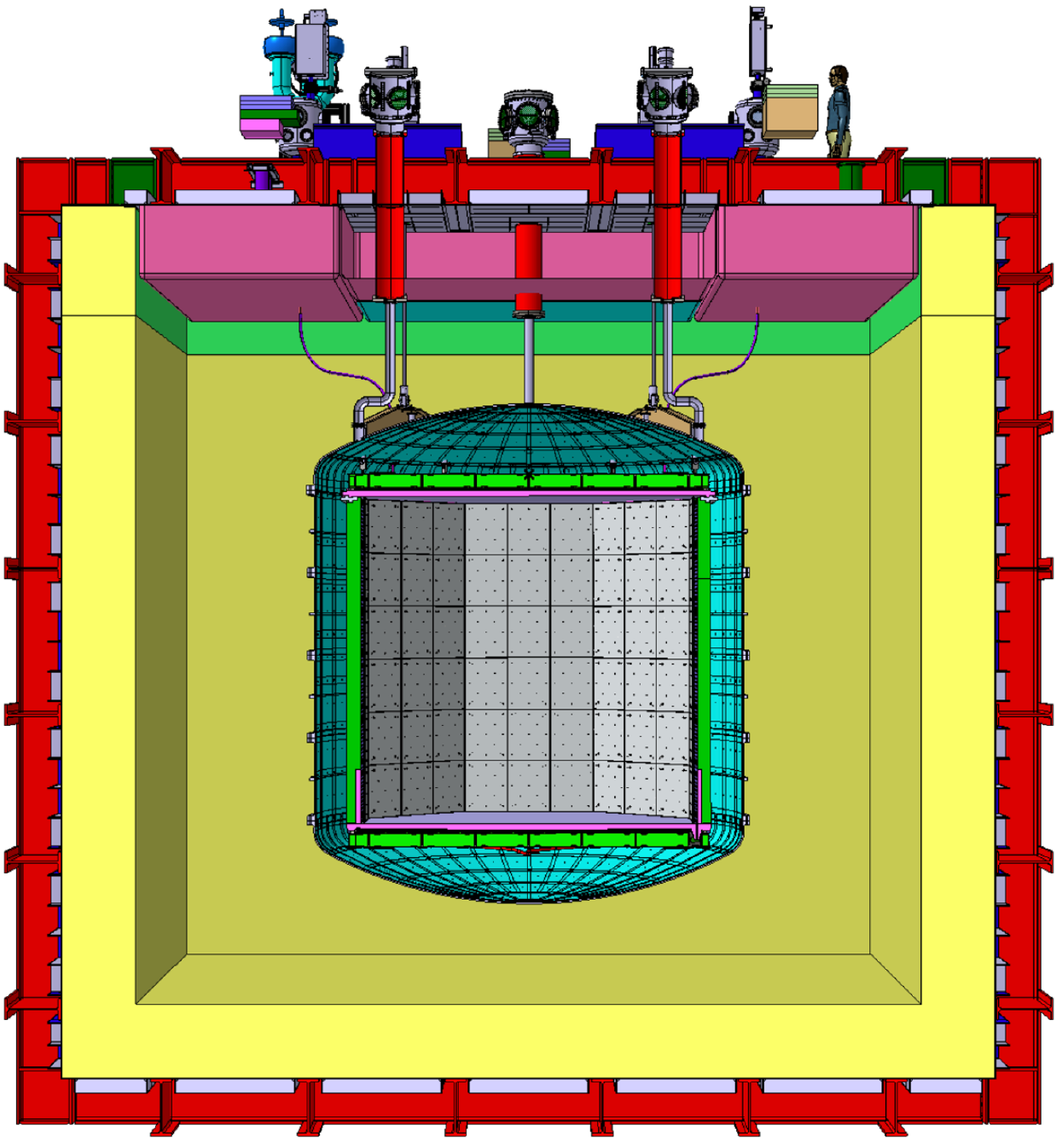}
    \label{fig:ds20k-section}
\end{subfigure}
\hspace{0.02\textwidth}
\begin{subfigure}{0.48\textwidth}
    \centering
    \includegraphics[width=\textwidth,height=5.5cm]{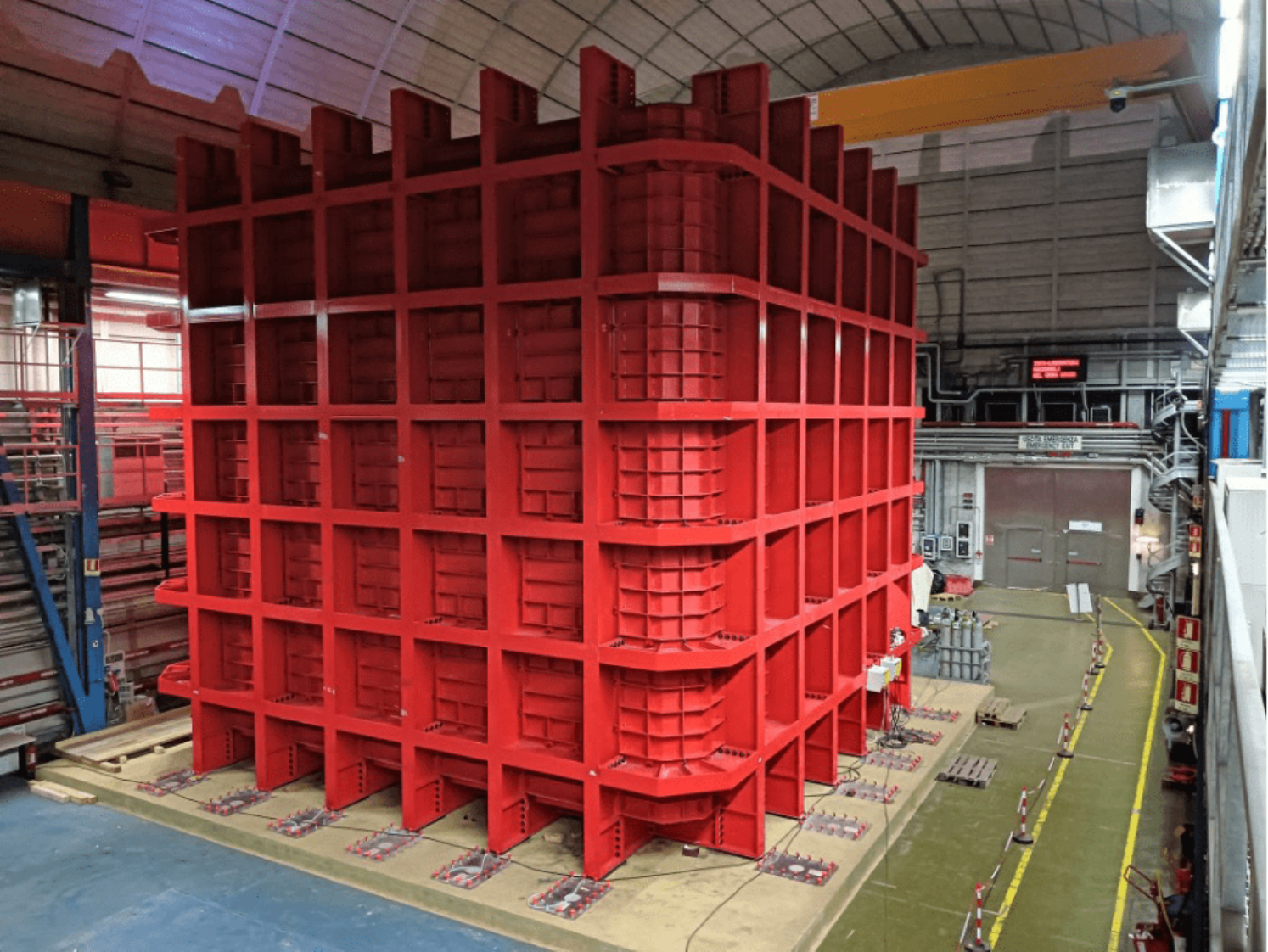}
    \label{fig:ds20k-picture}
\end{subfigure}
\caption{(Left) Vertical section of the DarkSide-20k detector with the ProtoDUNE-like membrane cryostat and the stainless steel vessel containing the TPC. (Right) Picture of the cryostat assembled at LNGS.}
\label{fig:ds20k-detector}
\end{figure*}

DS-20k is a massive liquid argon detector currently in construction at Laboratori Nazionali del Gran Sasso (LNGS), Italy, with leading sensitivity in the next decade in the DM mass range of \SI{1}{GeV/c^2} to \SI{10}{TeV/c^2}. The core of the detector is a double-phase time projection chamber (TPC) filled with \larvolume of liquid argon (\fiducialvolume fiducial), with a target instrumental background of < \(0.3\) events from neutrons and $\sim$ 3 events from neutrinos in the DM region of interest, in \(10\) years of operations.
The detector features a nested structure, formed by a ProtoDUNE-like membrane cryostat (Fig. \ref{fig:ds20k-detector}) \cite{Montanari:2015pxz}, enclosing the inner veto and the double-phase TPC. With a total volume of approximately \SI{580}{\meter^3} of atmospheric argon (AAr), the cryostat acts both as an outer muon veto and thermal bath. Both the inner veto and the TPC are immersed in low-radioactivity underground argon (UAr) and separated from the AAr by a stainless steel vessel.
The inner veto is filled with \SI{36}{t} of liquid argon and helps at efficiently tagging neutrons. The latter constitute the source of background, as they produce a DM-like signal in the TPC. The expected interaction of a DM particle inside the TPC is a coherent elastic scattering off an argon nucleus, which can be detected via a combination of the prompt scintillation light (S1 signal) and the delayed signal due to the drift of the ionization electrons produced by the recoiling nucleus (S2 signal) \cite{Agnes:2015gu}. 
Both signals are readout by Silicon Photomultiplier-based photodetectors located on the optical planes at the top and bottom of the TPC and provide energy information \cite{acerbi2024qualityassurancequalitycontrol}. The S2 signal allows \(x\)-\(y\) position reconstruction with a few cm resolution and the drift time between S1 and S2 signals is proportional to the \(z\) position, allowing for full tridimensional event reconstruction. 

The need for radiopure underground argon arises from the fact that atmospheric argon (AAr) is naturally radioactive due to the presence of the radioisotope \ce{ ^{39}Ar }, having an activity level of \SI{1}{\becquerel/\kilo\gram} \cite{BENETTI200783}. 
Given the low energy threshold of the TPC (< \SI{1}{\kilo e \volt ee}, where keVee refers to the amount of energy deposited by an electronic recoil) and the scale of its detection volume, this $\beta$-emitter can yield a too high probability of two or more signals overlapping in time, preventing an unambigous  association of S1 and S2 signals or even the deterioration of the PSD capability. On the other hand, argon extracted from underground sources, although it can be activated by cosmic rays during transportation, exhibits radioactivity reduced by more than a factor of one thousand, as demonstrated by the DarkSide-50 experiment \cite{Agnes_2016}. 


The DS-20k TPC is shaped as a regular octagonal prism, with a height and diameter of \SI{350}{cm} and formed by eight  Polymethyl Methacrylate (PMMA) panels. It is completed with two transparent acrylic windows on the bottom and the top, each coated with a thin conductive layer of Clevios. These coatings act as electrodes to generate a \SI{200}{\volt/\centi\meter} electrostatic field within the detector. This field allows the drift of the ionization electrons towards the top of the TPC, where a more intense field allows the extraction of the electrons into the thin layer of gaseous argon at the top of the chamber and the generation of UV scintillation light via electroluminescence forming eventually the S2 signal. A metallic grid, located a few millimetres beneath the gas layer separates the drift field from the extraction field region.
The walls of the TPC are coated with 98\% reflectivity reflector panels and TetraPhenyl Butadiene (TPB) wavelength shifter to convert the UV scintillation light of Ar to \SI{420}{\nano\meter} where the SiPMs have optimum performance.

\section{DarkSide-20k Data Acquisition System}
\label{sec:daq}
The Data Acquisition system (DAQ) for the DarkSide-20k experiment, is designed to continuously acquire signals from the \(2720\) photosensors of the TPC and the vetos. Individual readout channel waveforms are digitized and transferred to the following stage for further processing, without waiting for a trigger decision. In this sense, the DarkSide-20k experiment operates in \textit{triggerless mode}, where the data stream is uninterrupted and the isolation of interesting signals for physics searches is offloaded to an online computing farm (Sec. \ref{subsec:time-slice-architecture}). 
Digitized waveforms are transferred to external nodes called FrontEnd Processors (FEPs, Sec. \ref{subsec:front-end-processors}), where a matched filter and a peak finder are applied to identify the peaks, or \textit{hits},  within the waveform itself. Hits and associated information are time sorted and moved to dedicated nodes called Time Slice Processors (TSPs, Sec. \ref{subsec:time-slice-processors-merger}), where the online reconstruction is performed. Time slices from multiple TSPs are collected into the Merger machine (Sec. \ref{subsec:time-slice-processors-merger}), and stored on disk. Additionally, the Merger can conduct online processing of preselected data fragments for a combined physics analysis which can be used as an input to contribute to the Supernova Early Warning System (SNEWS 2.0) \cite{Al_Kharusi_2021}.

\subsection{Waveform digitization}
\label{subsec:waveform-digitization}
Waveforms from the entire detector will be digitized using 
\(48\) commercial VX2745 CAEN \SI{16}{\bit}, \SI{125}{\mega S\per\second}, \SI{4}{\volt} peak-to-peak, high channel density (\(64\) channels) waveform digitizers \cite{wfd5}. These modules support the integration of custom firmware, which can be uploaded to a reserved section of the FPGA using the OpenFPGA service \cite{openfpga}. This capability provides access to the raw digital data stream, enabling the implementation of custom acquisition control flows and data processing tasks such as triggering, data filtering, and compression before control is handed back to the CAEN firmware for data transmission via Ethernet to the FEPs.


The primary goal of the custom firmware is to implement a tailored algorithm for identifying segments of the digitized waveforms corresponding to signals as low as a single photo-electron with high efficiency. This process begins with the raw waveforms being processed through a  \(64\)-step (\SI{500}{\nano\second}) Finite Impulse Response (FIR) filter with user-defined shape (firmware parameter). This filter is needed to gain a factor \(2\) Signal to Noise Ratio (SNR), allowing to set a threshold of the order of \(0.5\) photoelectrons with negligible fake rate.
If the signal remains above the threshold for a minimum number of samples, the segment is retained, and the acquisition continues until the signal falls below a configurable secondary threshold.

\begin{figure}[ht!]
  \begin{center}
    \includegraphics[width=0.8\textwidth]{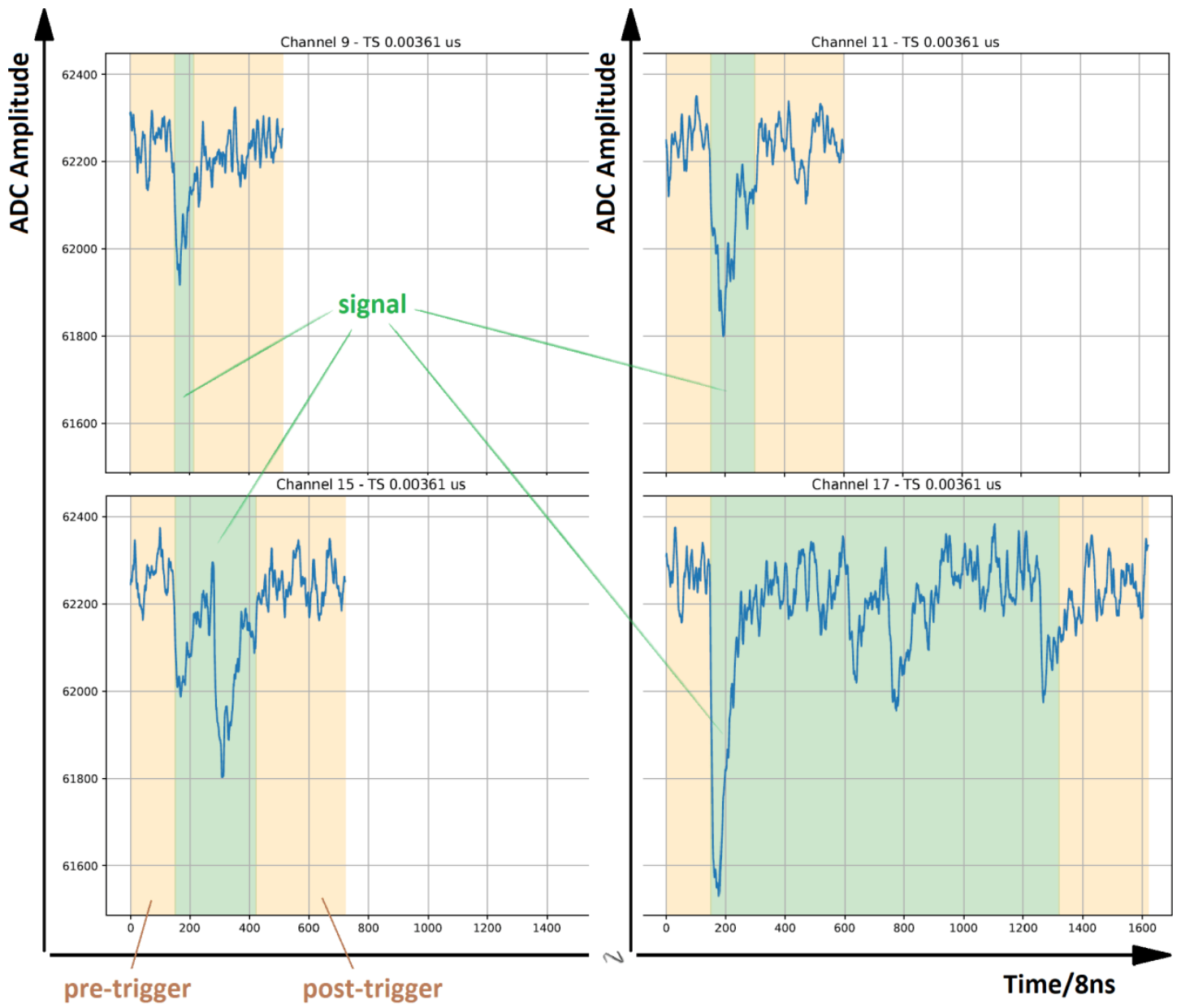}
  \end{center}
  \caption{Dynamic acquisition window in the CAEN VX2745 firmware: Filtered waveform segments exceeding a threshold are retained until the signal drops below another configurable threshold. The green region shows time-over-threshold, while yellow indicates pre- and post-trigger regions.}
\label{fig:ds20k-dynamic-window}       
\end{figure}

To ensure the extraction of essential quantities, such as the noise level, a small number of samples are appended to the end of the segment in the so-called post-trigger region, while a short pre-trigger region is included to capture baseline information.  
This \textit{gated acquisition} method optimizes digitizer throughput while preserving high peak-detection efficiency. If a subsequent peak triggers the threshold within the same acquisition, the gate is extended to include this new signal, along with its associated post-trigger region. 


\subsection{Time Slice Architecture}
\label{subsec:time-slice-architecture}

The DS-20k detector features \(2720\) readout channels, which must be processed simultaneously to enable efficient data filtering and reduction before final storage. The data rate generated by the TPC alone in terms of waveforms above the trigger threshold, defined at the digitizer level, is expected to reach approximately \tpcdatarate. This includes the Dark Count Rate (DCR), S1 signals and the dominating S2 signals. This estimation is based on the Monte Carlo background simulation with the current estimation of the radioactive contaminants in the detector materials and on a measured DCR of \dcr per channel. The desired data rate target after the acquisition chain is \SI{60}{\mega\byte/\second} for the TPC alone, corresponding to an online data reduction factor of about 50.

Merging the data from the full detector on a single processing unit while having sufficient latency to analyse data requires substantial computational resources.
To address this challenge, the data acquisition system employs the Time Slice approach. This method segments the acquisition timeline into discrete intervals, or "Time Slices", each of which is independently assigned to a Time Slice Processor (TSP) for further analysis. Once the processing of a Time Slice is complete, the TSP notifies an independent application, the Pool Manager (PM, Sec. \ref{subsec:pool-manager}), of its availability to handle a new Time Slice.  
Physics events spanning two consecutive Time Slices must be properly addressed, as each TSP operates without direct access to the preceding Time Slice. This challenge is mitigated by duplicating a fraction of the data from one Time Slice into the subsequent one. This overlapping segment ensures that boundary events occurring at the edges of Time Slices are fully captured and analyzed. The extent of the overlap corresponds to the maximum electron drift time in the Time Projection Chamber (TPC), approximately \SI{5}{\milli\second}.  
Given a default Time Slice duration of \SI{1}{\second}, this overlap results in a 0.5\% duplication of analyzed events within each overlapping window. This design choice is essential to guarantee continuity and completeness in event reconstruction across Time Slice boundaries.
A large number of channels implies that multiple digitizers are at work. Therefore the time segmentation mechanism requires the transmission of a Time Slice Marker (TSM) to all the digitizers to ensure a proper segment assembly based on the Time Slice number.

\subsection{FrontEnd Processors (FEPs)}
\label{subsec:front-end-processors}
\begin{figure*}[ht!]
\centering
\includegraphics[width=\textwidth]{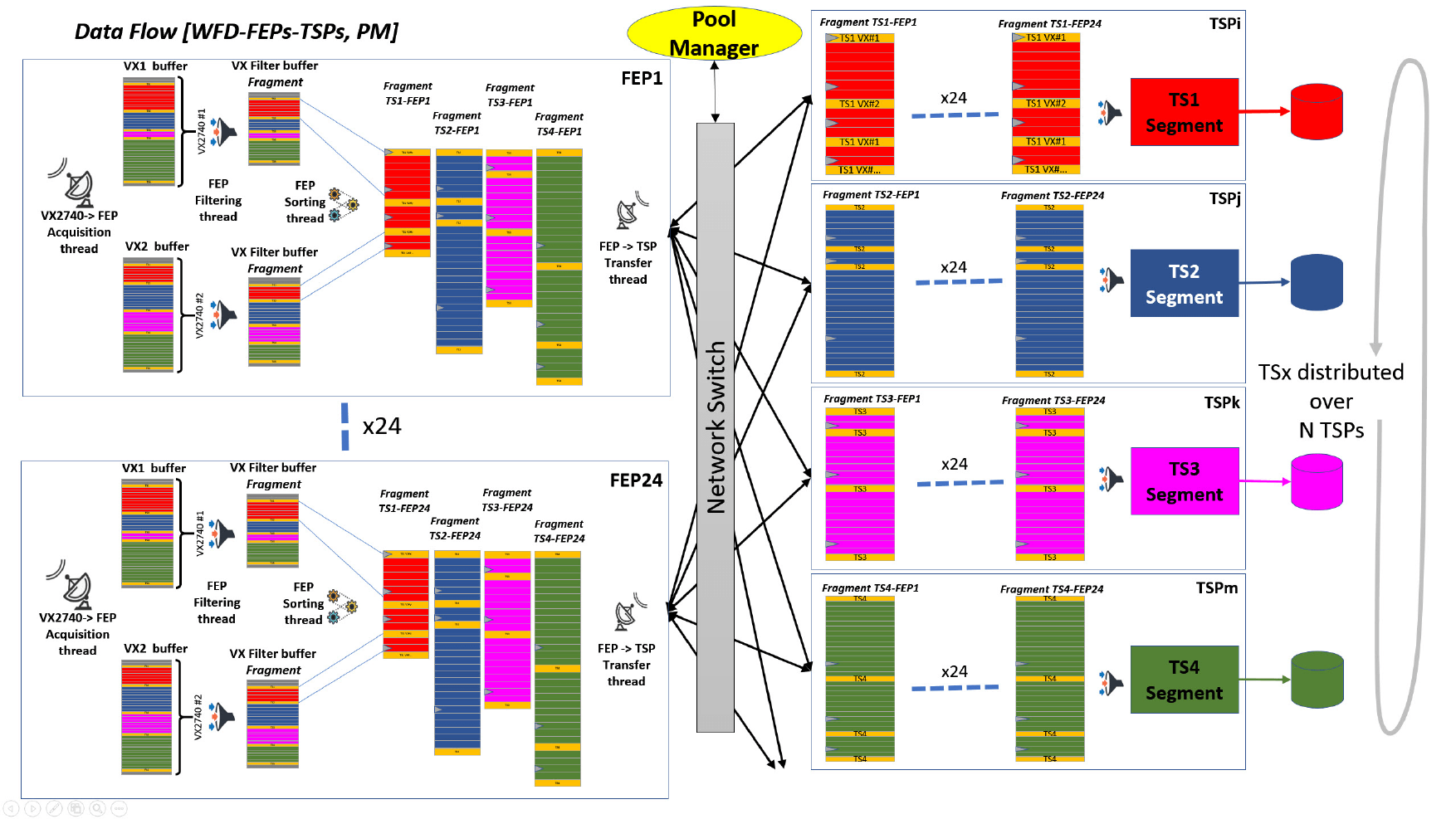}
\caption{Scheme of the DS-20k data flow: waveforms acquired by the WFDs are read out by the FEPs in \SI{1}{\second} Time Slices. Waveforms are then filtered, reduced to hit-level information and transferred to the TSPs for further processing. The management of the data fragment transmission to individual TSPs is left to the PM application.}
\label{fig:ds20k-data-flow}
\end{figure*}

The readout of the individual waveform digitizer (WFD) is performed by a dedicated processor (FEP). Due to power constraints, each FEP is capable of handling up to two WFDs. For the DS-20k detector, a total of 24 FEPs are employed to collect and process data from all waveform digitizers.
The management of the transmission of the data fragment to individual TSPs is left to the PM
application. Its role is to receive "idle" notification from any TSPs (once the previous Time Slice analysis has been completed) and broadcast to all the FEPs the destination address for the upcoming segment to the next available idle TSP (Sec. \ref{subsec:pool-manager}). Fig. \ref{fig:ds20k-data-flow} shows a pictorial representation of the Time Slice scheme and the time fragment dispatching mechanism.

The primary function of the FEPs in the DS-20k data acquisition system is to read out, perform an initial data reduction, and sort data segments received from the WFDs corresponding to predefined Time Slices. Additionally, the FEPs are responsible for extracting the timing and amplitude of all peaks, or "hits", within these segments.
Waveforms are filtered using an Auto-Recursive (AR) matched filter designed to match the SiPM response shape. By relying solely on the current input and the previous output, it minimizes computational complexity, enabling real-time processing of high-throughput data.
The hit-finding algorithm identifies hits based on their prominence, calculated as the difference between the filtered waveform and a smoothed version of the waveform obtained using a low-bandwidth low-pass filter. This approach enhances the efficiency of detecting multiple photoelectrons that are closely spaced in time within a single waveform segment.
To further suppress fake hits due to noise, only signals with a charge-to-prominence ratio greater than a specified minimum value are retained. The hit-finding algorithm is governed by three key parameters: the window length for the moving average, the signal threshold, and the charge-to-prominence ratio. 

The FEPs are also responsible for monitoring the individual channels and assessing the quality of the raw data. This monitoring functionality enables data extraction at the FEP level, which can then be transmitted to a remote Data Monitoring Computer (DMC) via the MIDAS event transport, either periodically or on request.  
Channel-level monitoring is conducted to maintain data integrity throughout the acquisition process. Care is taken to ensure that this monitoring has minimal impact on the extraction of physics information, preventing any interference with the primary data stream.

\subsection{Data Flow Control: Pool Manager (PM)}
\label{subsec:pool-manager}
The data transfer from the FEPs to the TSPs is orchestrated by the PM application, which operates on the MIDAS server. Its primary function is to assign the appropriate TSP address to each FEP, enabling the transmission of Time Slice data. Once a TSP has completed the analysis of its assigned Time Slice, it is responsible for notifying the PM. This notification includes not only the completion status of the analysis but also a detailed report on the outcome of the last TSM analysis.


Upon receiving the notification packet, the PM processes it in two distinct steps. First, the PM adds the TSP-ID to an idle processor queue, making the TSP available for handling the next Time Slice. Second, the PM composes an event status for the recently completed Time Slice. This status, along with related information, is integrated into the MIDAS software infrastructure and can be accessed via a custom web history page, allowing for real-time monitoring and review of processing events.
The PM continuously scans the Idle processor queue (FIFO), it retrieves the oldest TSP-ID and pairs it with the next available TimeSlice-ID, forwarding this information to all the FEPs. Each FEP then stores the TSP-ID and corresponding TimeSlice-ID in its queue, allowing multiple Time Slice data segments to be simultaneously transferred to different TSP destinations. This parallel processing scheme significantly optimizes network bandwidth utilization. Communication between the TSP, PM, and FEPs occurs through the main data switch using ZeroMQ (ZMQ) message packets, which provide an efficient method for message broadcasting \cite{zmq}. Despite sharing the main data network, the communication overhead introduced by ZMQ is minimal and has negligible impact on the overall data transfer performance.

The overall Data Acquisition system is managed by the MIDAS software package \cite{842578,10126097,5750358}, which performs several key functions: configuring the readout equipment, orchestrating the run sequence, generating various alarm levels (e.g., warnings, errors, or custom alarms) based on user-defined criteria, controlling data transfer to analysis tools, recording data to permanent storage, and managing equipment operation and monitoring. Additionally, MIDAS provides an application framework for device interfaces.

In the DS-20K experiment architecture, while MIDAS continues to handle overall control and monitoring, its role has been specifically adapted to manage data flow control only, rather than data transfer. This architectural decision enables the use of raw socket data transfer from the Digitizers to the TSPs, allowing for optimal data throughput in the system.

\subsection{Time Slice Processors (TSPs) and Merger}
\label{subsec:time-slice-processors-merger}

The TSPs are responsible for managing incoming connections from the FEPs and receiving their data payloads. Once a TSP completes the analysis of a Time Slice, it notifies the PM and then waits for the next data fragment to arrive.
The primary role of the TSPs is to handle the collection and selection of hits to ensure efficient data processing. However, the TSPs' functionality can be expanded to perform more sophisticated analyses, including event classification. Events can be categorized into various types such as:
Regular DM (e.g. S1 in the WIMP region of interest), high energy gammas in the TPC for calibration (e.g. high energy S1), low energy S2 events (e.g. low multiplicity S2 pulses isolated from other pulses), and Inner/Outer Veto (e.g., pulse above the threshold necessary for cosmogenic suppression and calibration or monitoring).   
This classification process enables the system to pre-scale certain event types, such as high-energy gammas or alphas, which can reduce the overall demand for data storage. By performing these analyses, the TSPs generate a new set of data that is stored locally on each TSP's storage device.
TSPs notify the PM at key stages of their operations: when the analysis of a Time Slice begins, if the analysis fails, and when the transmission of the Time Slice to the next stage of data acquisition is completed. The PM then relays this information to the Merger, a secondary machine responsible for collecting the Time Slices. 

The Merger sorts the Time Slices chronologically and concatenates them into a continuous data stream. Once the Time Slices are properly organized, the Merger stores them on a local disk for temporary storage. Additionally, it handles the transfer of the concatenated Time Slices to the CNAF data centre for long-term storage and offline analysis.
Moreover, the Merger continuously monitors for missing Time Slices during the collection process. 
Any Time Slices that are not immediately delivered to the Merger are temporarily stored in the corresponding TSP until they can be successfully transmitted. This mechanism ensures that the system can handle delays or failures in the data acquisition process without losing data.

In addition to sorting and concatenating Time Slices, the Merger can perform a fast supernova trigger by analyzing Time Slices near those flagged by the TSPs, where the event rate exceeds the baseline beyond specific energy ranges.
If a potential supernova event is detected the Merger will trigger an alert. This alert is then communicated to the Supernova Early Warning System (SNEWS 2.0) \cite{Al_Kharusi_2021}, a global network designed to provide early notification of supernovae by gathering signals from neutrino detectors around the world.

\section{Summary}
\label{sec:summary}

The DarkSide-20k experiment employs a sophisticated triggerless Data Acquisition (DAQ) system optimized for the high data rates expected for such a massive detector. The DAQ digitizes signals from 2720 channels using high-performance waveform digitizers, processes data in 1-second Time Slices, and uses Front-End Processors (FEPs) to extract the photoelectron content of the waveforms. Time Slice Processors (TSPs) perform real-time data reconstruction and classification, while the Merger accumulates Time Slices, manages their storage, and facilitates rapid analyses such as supernova alerts.  
This modular architecture ensures scalability, efficient bandwidth use, and minimal data loss, with an emphasis on robust data quality monitoring and high sensitivity. 

\bibliography{template.bib}

\end{document}